\documentclass[letterpaper,12pt]{article}
\usepackage{graphicx}
\usepackage{dcolumn}
\usepackage{bm}
\usepackage{osajnl}
\usepackage[draft]{hyperref}

\begin{document}

\title{Spatial mode properties of plasmon assisted transmission}
\author{Xi-Feng Ren,Guo-Ping Guo\footnote{gpguo@ustc.edu.cn},Yun-Feng Huang,Zhi-Wei Wang,Guang-Can Guo}
\address{Key Laboratory of Quantum Information and Department
of Physics, University of Science and Technology of China, Hefei
230026, People's Republic of China}

\begin{abstract}
Orbital angular momentum of photons is explored to study the
spatial mode properties of plasmon assisted transmission process.
We found that photons carrying different orbital angular momentums
have different transmission efficiencies, while the coherence
between these spatial modes can be preserved.
\end{abstract}
\ocis{230.3990, 240.6680, 030.1640.}
\maketitle

It has long been observed that there is an unusually high optical
transmission efficiency in metal films perforated with a periodic
array of subwavelength apertures\cite{1}. Generally, it is believed
that metal surface plays a crucial role and the phenomenon is
mediated by surface plasmons (SPs) and there is a process of
transform photon to surface plasmon and back to
photon\cite{4,crucial,ebbesen5}. In 2002, E. Altewischer \textit{et
al.} \cite{alt} first addressed the question of whether the
entanglement survives in this extraordinary enhancement light
transmission. They showed that quantum entanglement of photon pairs
can be preserved when they respectively travel through a hole array.
Therefore, the macroscopic surface plasmon polarizations, a
collective excitation wave involving typically $10^{10}$ free
electrons propagating at the surface of conducting matter, have a
true quantum nature.

\begin{figure}[b]
\centering
\includegraphics[width=7.0cm]{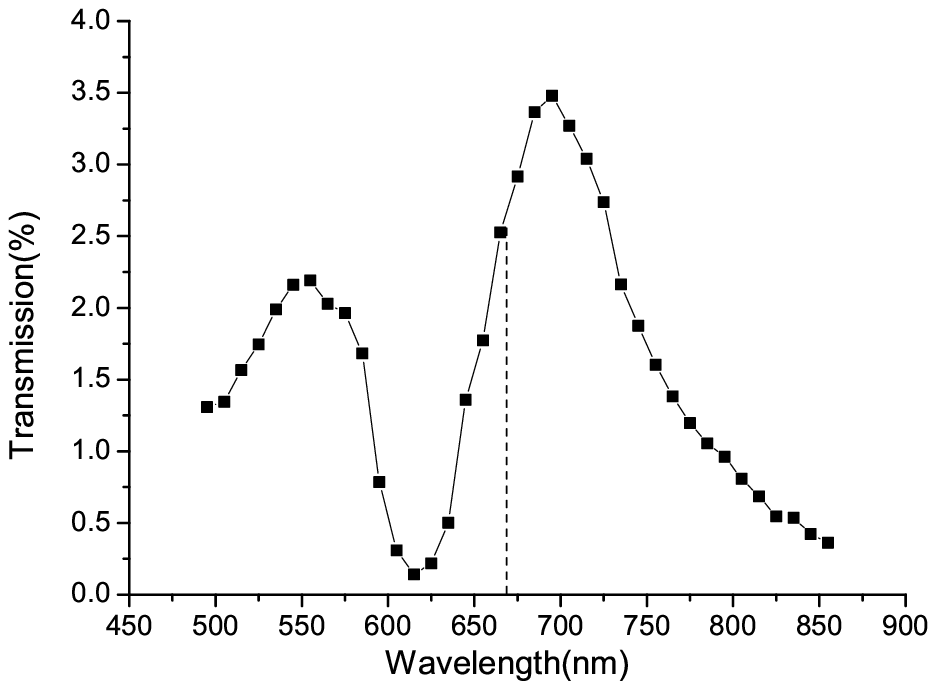}
\caption{Hole array transmittance as a function of wavelength. The dashed
vertical line indicates the wavelength of 670nm used in the experiment.}
\end{figure}

\begin{figure}[b]
\centering
\includegraphics[width=8.0cm]{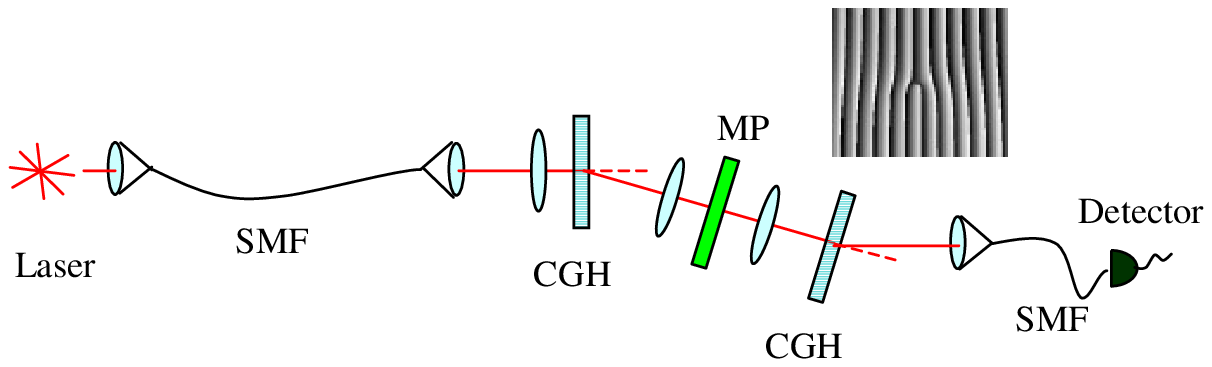}
\caption{Experimental set-up. A computer generated hologram(CGH)
is used to change OAM  of the light from output of a single-mode
fiber(SMF). The metal plate(MP) is placed between the twin
lenses(focus 35mm). The light incident on the metal plate has a
diameter small than $30\mu m$. Another CGH and SMF are combined to
analyze the OAM of the transmitted photons. Inset, picture of part
of a typical CGH($l=1 $).}
\end{figure}

So far, polarization properties of nanohole arrays have been studied
in many works\cite{Elli04,Gordon04,Altew05}. Then, how about the
spatial mode properties of nanohole arrays in the photon to plasmon
and back to photon process? The metal plates used in previous
experiments always have special structure-usually like lattice.
Modes of surface plasmons are correlated with these typical
structures. If the light incidented on the metal plate has a
transverse spatial distribution itself, what will happen? What are
the spatial modes of the transmitted photons? Understanding these
questions will help us to have an in-depth understanding on the
plasmon-assisted transmission process. In our experiment, photons
with different orbital angular momentums (OAM) are incidented on the
metal plate and OAM of transmitted photons of zero-order are
investigated. We find that for photons carrying integral OAM, the
spatial modes are not changed, while different OAM corresponding to
different transmission efficiencies. The coherence between different
spatial modes can also be preserved.

Fig. 1 shows the metal plate transmittance as a function of
wavelength. The dashed vertical line indicates the wavelength of
$670$ $nm$ used in our experiment. The transmission efficiency of
the metal plate at $670nm$ is about $2.5\%$, which is much larger
than the value of $0.55\%$ obtained from classical theory\cite
{Bethe}. The metal plate is produced as follows: after subsequently
evaporating a $3$-$nm$ titanium bonding layer and a $135$-$nm$ gold
layer onto a $0.5$-$mm$-thick silica glass substrate, a Focused Ion
Beam Etching system (FIB, DB235 of FEB Co.) is used to produce
cylindrical holes ($200nm$ diameter) arranged as a square lattice
($600nm$ period). The total area of the hole array is $30\mu m\times
30\mu m$, which is actually made up of four hole arrays of $15\mu
m\times 15\mu m$ area for the technical reason.

It has been shown that paraxial Laguerre-Gaussian(LG) laser beams
carry a well-defined OAM\cite{Allen92}, and LG modes form a complete
Hilbert space. If the mode function is a pure LG mode with winding
number $l$ , then every photon of this beam carries an OAM of
$l\hbar $. This corresponds to an eigenstate of the OAM operator
with eigenvalue $l\hbar $\cite{Allen92}. If the mode function is not
a pure LG mode, each photon of this light is in a superposition
state, with the weights dictated by the contributions of the
comprised different $l$th angular harmonics.

Usually, we use computer generated holograms
(CGHs)\cite{ArltJMO,VaziriJOB,Ren04} to change the winding number of
LG mode light. It is a kind of transmission holograms. Corresponding
to the diffraction order $m$, the hologram can change the winding
number of the input beam by $\Delta l_m=ml$. The diffraction
efficiency depends on the phase modulation $\delta $. In our
experiment, the efficiencies of CGHs are all about $30\%$. Fig. 2.
shows part of a typical CGH($l=1$) with a fork in the center.
Generally, it is also important to be able to both produce and
analyze superposition states in a chosen basis. A convenient method
for creating superposition modes is to use a displaced
hologram\cite{VaziriJOB}, which is particularly suitable for
producing superpositions of an $LG_{0l}$ mode with the Gaussian
mode. The Gaussian mode ($l=0$) can be identified using single-mode
fibers in connection with avalanche detectors. All other modes have
a larger spatial extension, and therefore cannot be coupled into the
single-mode fiber efficiently. High order LG modes ($l\neq 0$) are
identified using mode detectors consisting of CGH and single-mode
optical fiber(Shown in Fig. 2.). This mode detector can also be used
to identify the superposition mode by displace the
CGH\cite{VaziriJOB,Langford04}.

\begin{figure}[b]
\centering
\includegraphics[width=4.0cm,height=5.0cm]{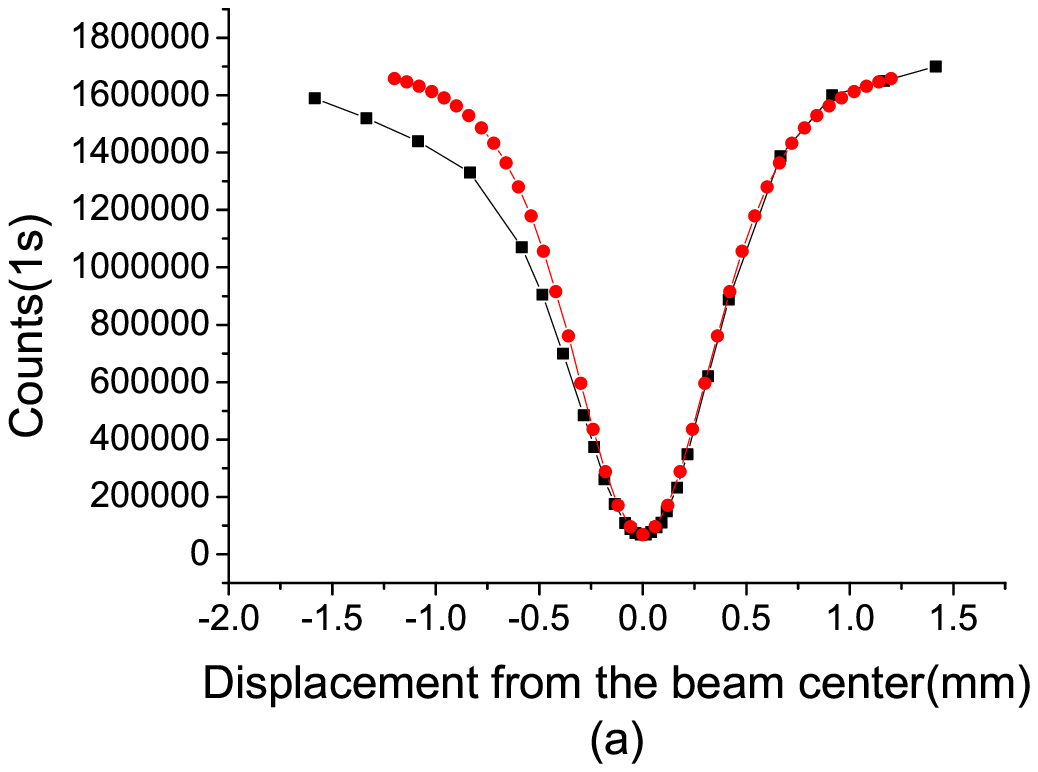} %
\includegraphics[width=4.0cm,height=5.0cm]{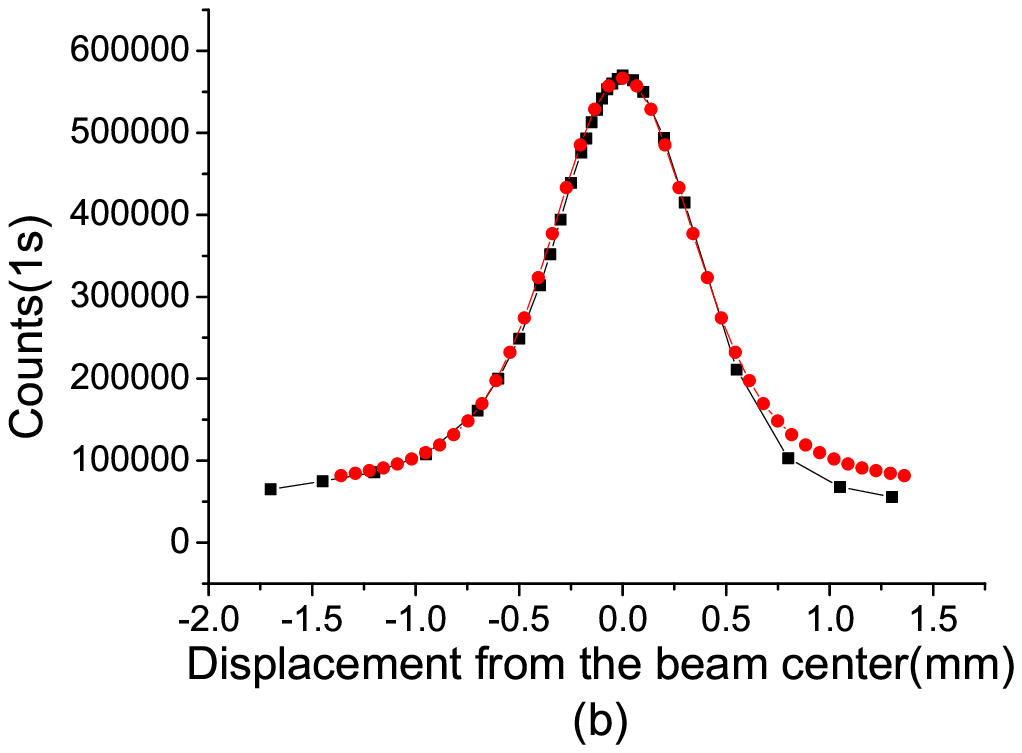}
\caption{Counts(square dots) as a function of the displacement of
the second CGH. The metal plate is moved out. (a) The fork of first
CGH is displaced far away from the beam center, so the light is in
$\left| 0\right\rangle$ OAM state. (b) The fork of first CGH is
placed in the beam center, so the light is in $\left|
1\right\rangle$ OAM state. The round dots come from theoretical
calculation.}
\end{figure}

\begin{figure}[b]
\centering
\includegraphics[width=4.0cm,height=5.0cm]{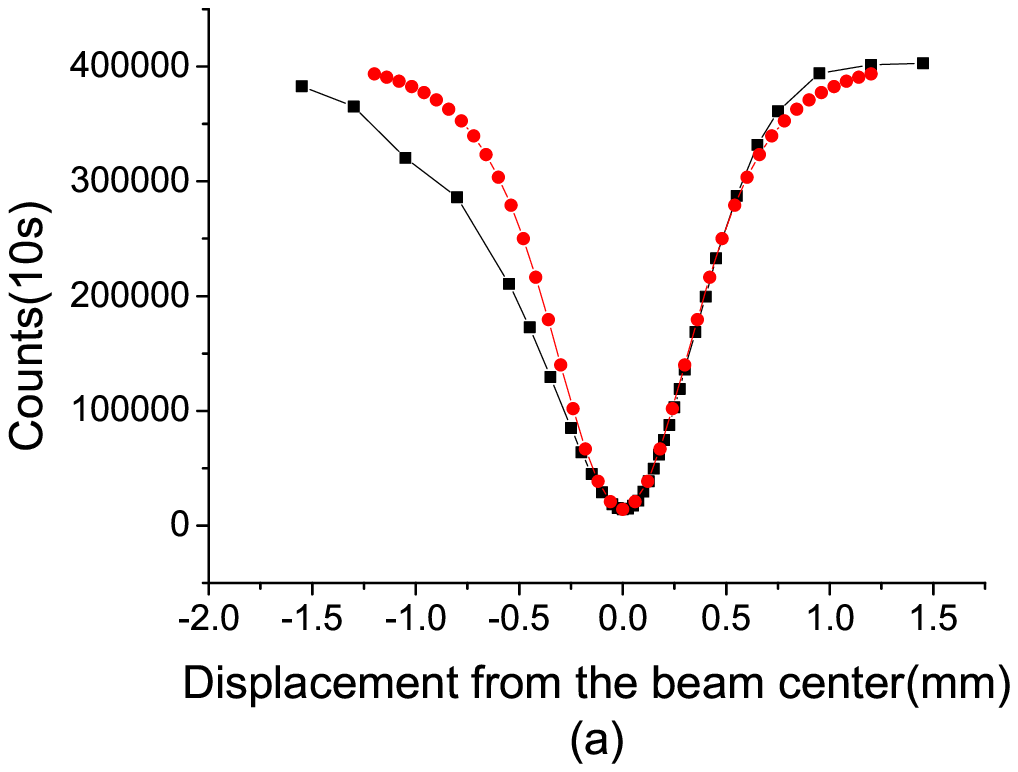} %
\includegraphics[width=4.0cm,height=5.0cm]{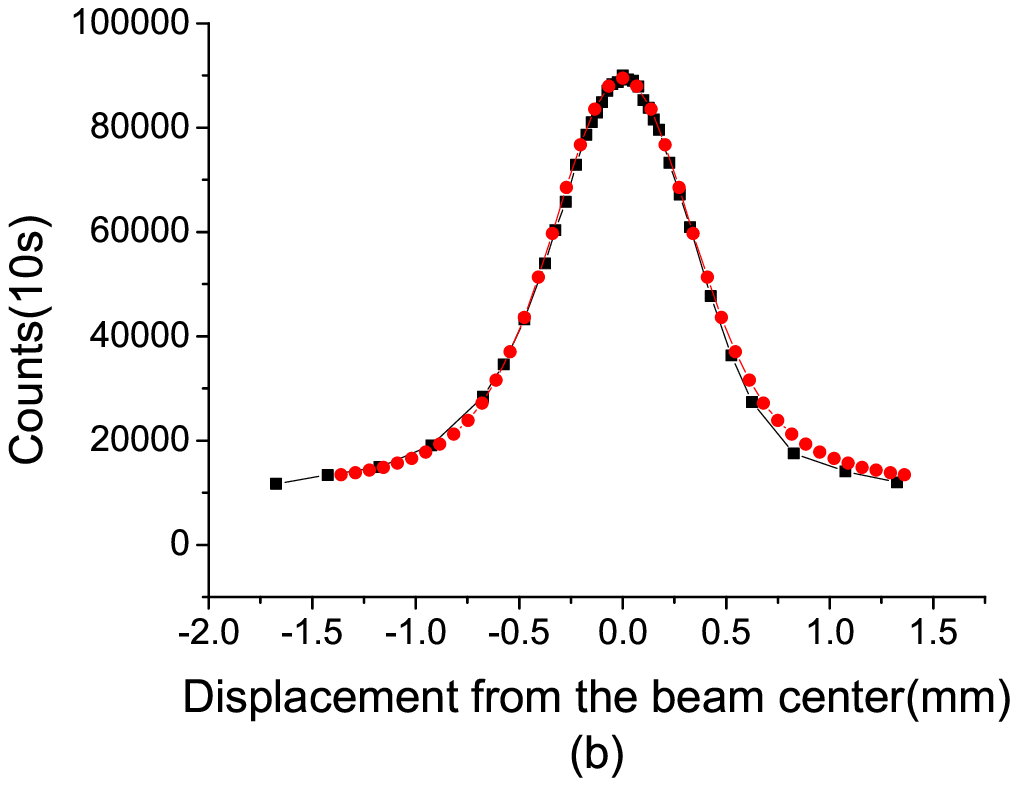}
\caption{Counts(square dots) as a function of the displacement of
the second CGH. The metal plate is placed between the twin-lenses.
(a) The fork of first CGH is displaced far away from the beam
center, so the light incident on the metal plate is in $\left|
0\right\rangle$ OAM state. (b) The fork of first CGH is placed in
the beam center, so the light incident on the metal plate is in
$\left| 1\right\rangle$ OAM state. The round dots come from
theoretical calculation.}
\end{figure}

The experimental setup is shown in Fig. 2. The light is in
vertical polarization with $670nm$ wavelength. A CGH ($\Delta
l=1$) is used to change OAM of the light from output of a
single-mode fiber. The metal plate is placed between the twin
lenses(focus 35mm), so the light incidented on the metal plate has
a diameter about $25\mu m$. Another CGH ($\Delta l=-1$) and
single-mode fiber are combined to analyze the OAM of the
transmitted photons. Silicon avalanche photodiode (APD) photon
counter is used to record counts(We use neutral density filter to
reduce the intensity of transmitted zero-order light). First of
all, we remove the metal plate from the twin-lenses. When the fork
of first CGH is placed far away form the beam center, OAM of the
light is not changed and the photons are still in OAM state
$\left| 0\right\rangle$. In this case, we move the second CGH from
one side to another side horizontally and record the counts. Fig.
3(a) shows the experiment result, which fits nicely with the
theoretical numerical calculation. Then the first CGH is placed
with the fork in beam center, which changes the OAM state of
photons to $\left| 1\right\rangle$. We also move the second CGH
form one side to another side horizontally and record the counts.
The experiment result is shown in Fig. 3(b), which also fits
nicely with the numerical calculation. These results show that the
shifted holograms work as described above.

Now we put the metal plate with hole array between the twin
lenses. Firstly, We move the second CGH horizontally and record
the counts with photons of Gaussian mode incidented on the metal
plate. As shown in Fig. 4(a), the counts have a similar curve
which indicates the spatial mode of the light is not changed, OAM
state of the transmitted photon is also $\left| 0\right\rangle$.
For the cases that input photons in OAM states $\left|
1\right\rangle$ or $\left| -1\right\rangle$(not shown in Fig. 4.),
the spatial modes are also preserved. So the spatial modes are not
changed for photons carrying integral OAM. It is also found that
for different OAM modes light, the transmission efficiencies are
different. For light in $\left| 0\right\rangle$ OAM state, the
transmission efficiency is $2.27\pm 0.27$\%; while for light in
$\left| 1\right\rangle$($\left|- 1\right\rangle$) OAM state, the
transmission efficiency is $1.56\pm 0.02 $\%($1.42\pm 0.10$\%).

\begin{figure}[b]
\centering
\includegraphics[width=4.0cm,height=5.0cm]{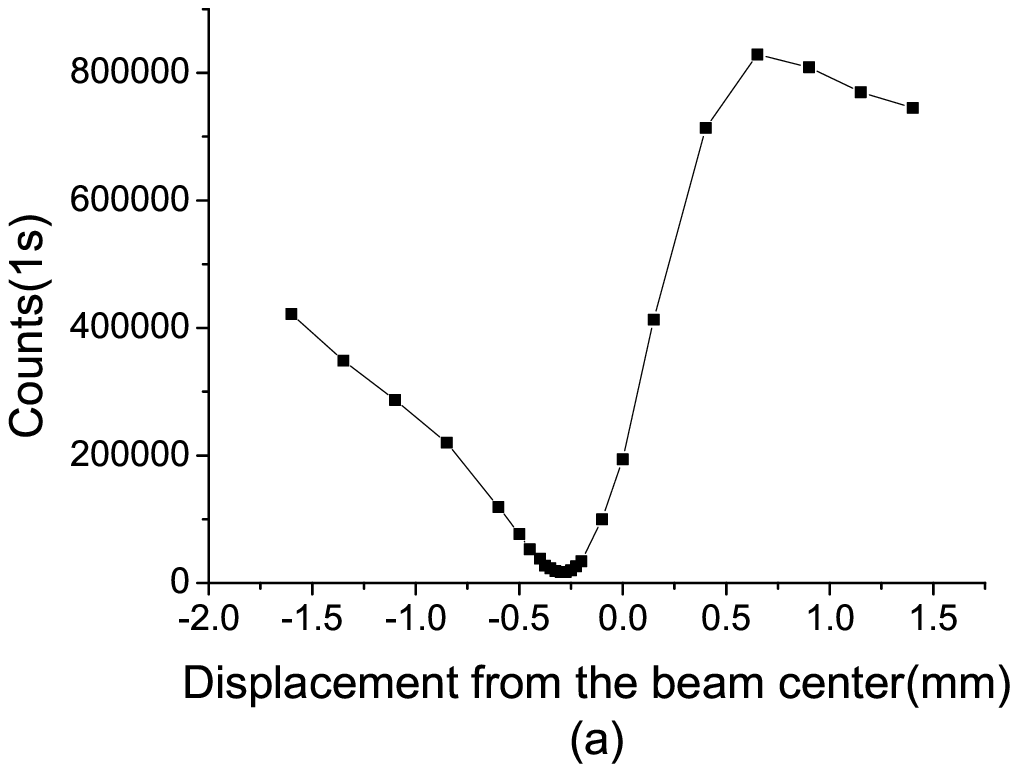} %
\includegraphics[width=4.0cm,height=5.0cm]{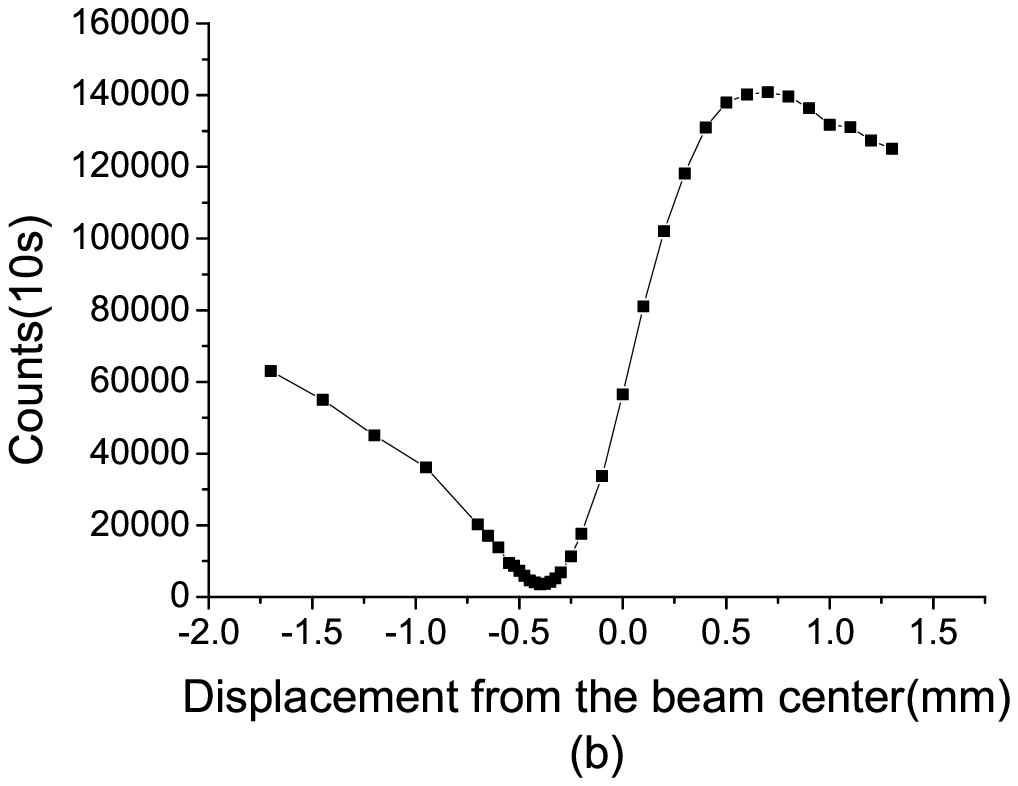}
\caption{Counts as a function of the displacement of the second CGH.
The fork of first CGH is displaced a small mount from the beam
center, so the light is in superposition mode.(a) The metal plate is
moved out from the twin-lenses. The visibility is 96.0\%.(b) The
metal plate is placed between the twin-lenses. The visibility is
94.4\%.}
\end{figure}

To investigate the decoherence between different modes in the photon
to plasmon and back to photon process, we need to input
superposition mode light on the metal plate. In the experiment, it
is done by displacing the fork of first CGH from the beam center.
The incident OAM state on the metal plate is $(a\left|
0\right\rangle +b\left| 1\right\rangle )/\sqrt{a^2+b^2}$, where $a$
and $b$ are real numbers. In the case of no metal plate, if we move
the second hologram from one side to another side horizontally,
there must be a special place that we have no count. Because in this
special place the second hologram change the state of input light
from $(a\left| 0\right\rangle +b\left| 1\right\rangle
)/\sqrt{a^2+b^2}$ to $(m\left| 1\right\rangle+n\left|
-1\right\rangle)/\sqrt{m^2+n^2}$, which can not be transmitted by
single-mode fiber. The counts are shown in Fig. 5(a) with visibility
$96.0\%$. Then we place the metal plate between the twin-lenses. If
there is decoherence between different OAM states ( here $\left|
0\right\rangle$ and $\left| 1\right\rangle$), there should be a
random phase $\varphi (t)$ between $\left| 0\right\rangle$ and
$\left| 1\right\rangle$ of output photons from metal plate(This
random phase actually corresponds to a dephasing process). The state
will change to $(a^{\prime }\left| 0\right\rangle +e^{i\varphi
(t)}b^{\prime }\left| 1\right\rangle )/\sqrt{a^{\prime
}{}^2+b^{\prime }{}^2} $, where $a^{\prime }$ and $b^{\prime }$ are
real numbers. Then however we move the second CGH, we can always
have counts for counts in experiment corresponding to sum of coming
photons in a period of time. On the contrary, if there is no
decoherence (or dephasing), we can also find a position of the
second CGH where there is no count. The experimental result is shown
in Fig. 5(b). We still find the no-count place, and the visibility
is $94.4\%$. There is a small distance between this no-count place
and the no-count place with no metal plate (Shown in Fig. 5), which
is coming from the different transmission efficiencies of different
modes light. The coherence between state $\left| 0\right\rangle$ and
state $\left| -1\right\rangle$ is also investigated by changing the
CGHs. The visibilities without and with the metal plate are $81.9\%$
and $82.0\%$ respectively. So we can say, the coherence between
different modes is preserved and there is no dephasing between
different OAM states in this plasmon assisted transmission.

In conclusion, we have investigated the spatial mode properties of
our typical nanohole arrays in the photon to plasmon and back to
photon process. We found that for photons carrying integral OAM,
the spatial mode is not changed, while the transmission
efficiencies for different modes light are different. The
efficiencies are concerned with the spatial distribution of the
input light. A more interesting phenomenon is that the coherence
between different modes is preserved. These results may give us
more hints to the understanding of plasmon-assisted transmission
of photons.

\begin{center}
\textbf{Acknowledgments}
\end{center}

This work was funded by the National Fundamental Research Program
(2001CB309300), National Nature Science Foundation of China
(10304017), the Innovation Funds from Chinese Academy of Sciences
and the Program of the Education Department of Anhui Province (Grant
No.2006kj074A).

\begin{References}
\bibitem{1}  T.W. Ebbesen, H. J. Lezec, H. F. Ghaemi, T. Thio, and P. A. Wolff,  Nature 391, 667 (1998).

\bibitem{4}  H. Raether, {\it Surface Plasmons on Smooth and Rough Surfaces
and on Gratings, Springer Tracts in Modern Physics, } Springer, Berlin, 1988
Vol. 111.

\bibitem{crucial}  D. E. Grupp, H. J. Lezec, T. W. Ebbesen, K. M. Pellerin, and Tineke Thio, Appl. Phys. Lett. 77 1569
(2000).

\bibitem{ebbesen5}  M. Moreno, F. J. Garc¨ªa-Vidal, H. J. Lezec, K. M. Pellerin, T. Thio, J. B. Pendry, and T. W. Ebbesen, Phys. Rev. Lett. 86, 1114
(2001).

\bibitem{alt}  E. Altewischer, M. P. van Exter and J. P. Woerdman, Nature 418 304 (2002).

\bibitem{Elli04} J. Elliott, I. I. Smolyaninov, N. I. Zheludev, and A. V. Zayats, Opt. Lett. 29, 1414 (2004).

\bibitem{Gordon04} R. Gordon, A. G. Brolo, A. McKinnon, A. Rajora, B. Leathem, and K. L. Kavanagh, Phys. Rev. Lett. 92, 037401
(2004).

\bibitem{Altew05} E. Altewischer, C. Genet, M. P. van Exter, and J. P. Woerdman, Opt. Lett. 30, 90 (2005).

\bibitem{Bethe}  H. A. Bethe, Phys. Rev. 66, 182 (1944).

\bibitem{Allen92} L. Allen, W. BeijersbergenM, R. J. C. Spreeuw and J. P. Woerdman, Phys. Rev. A 45
8185 (1992).

\bibitem{ArltJMO} J. Arlt, K. Dholakia, L. Allen and M. J. Padgett, J. Mod. Opt. 45
1231 (1998).

\bibitem{VaziriJOB} A. Vaziri, G. Weihs, A. Zeilinger, J. Opt. B:
Quantum Semiclass. Opt 4 s47 (2002).

\bibitem{Ren04} X. F. Ren, G. P. Guo, B. Yu, J. Li, and G. C. Guo,J. Opt. B: Quantum Semiclass. Opt 6 243
(2004).

\bibitem{Langford04}  N. K. Langford, R. B. Dalton, M. D. Harvey, J. L.
O'Brien, G. J. Pryde, A. Gilchrist, S. D. Bartlett, and A. G. White,
Phys.Rev.Lett. {\bf 93}, 053601, (2004).

\end{References}
\newpage

\section*{List of Figure Captions}

Fig. 1. Hole array transmittance as a function of wavelength. The
dashed vertical line indicates the wavelength of 670nm used in the
experiment.

Fig. 2. Experimental set-up. A computer generated hologram(CGH) is
used to change OAM  of the light from output of a single-mode
fiber(SMF). The metal plate(MP) is placed between the twin
lenses(focus 35mm). The light incident on the metal plate has a
diameter small than $30\mu m$. Another CGH and SMF are combined to
analyze the OAM of the transmitted photons. Inset, picture of part
of a typical CGH($l=1 $).

Fig. 3. Counts(square dots) as a function of the displacement of the
second CGH. The metal plate is moved out. (a) The fork of first CGH
is displaced far away from the beam center, so the light is in
$\left| 0\right\rangle$ OAM state. (b) The fork of first CGH is
placed in the beam center, so the light is in $\left|
1\right\rangle$ OAM state. The round dots come from theoretical
calculation.

Fig. 4. Counts(square dots) as a function of the displacement of the
second CGH. The metal plate is placed between the twin-lenses. (a)
The fork of first CGH is displaced far away from the beam center, so
the light incident on the metal plate is in $\left| 0\right\rangle$
OAM state. (b) The fork of first CGH is placed in the beam center,
so the light incident on the metal plate is in $\left|
1\right\rangle$ OAM state. The round dots come from theoretical
calculation.

Fig. 5. Counts as a function of the displacement of the second CGH.
The fork of first CGH is displaced a small mount from the beam
center, so the light is in superposition mode.(a) The metal plate is
moved out from the twin-lenses. The visibility is 96.0\%.(b) The
metal plate is placed between the twin-lenses. The visibility is
94.4\%.

\end{document}